\documentclass[sigconf]{acmart}

\AtBeginDocument{%
  }

\copyrightyear{2025}
\acmYear{2025}
\setcopyright{rightsretained}
\acmConference[RecSys '25]{Proceedings of the Nineteenth ACM Conference on Recommender Systems}{September 22--26, 2025}{Prague, Czech Republic}
\acmBooktitle{Proceedings of the Nineteenth ACM Conference on Recommender Systems (RecSys '25), September 22--26, 2025, Prague, Czech Republic}\acmDOI{10.1145/3705328.3748103}
\acmISBN{979-8-4007-1364-4/2025/09}

\begin{document}

%%
%% The "title" command has an optional parameter,
%% allowing the author to define a "short title" to be used in page he
%\title{LENS: Scaling Nuance Understanding for LLM-Powered Video RecSys}
\title{LLM-Powered Nuanced Video Attribute Annotation for Enhanced Recommendations}

%%
%% The "author" command and its associated commands are used to define
%% the authors and their affiliations.
%% Of note is the shared affiliation of the first two authors, and the
%% "authornote" and "authornotemark" commands
%% used to denote shared contribution to the research.
\author{Boyuan Long}
\authornote{Equal Contribution}
\affiliation{%
  \institution{Google}
  \city{Mountain View}
  \state{CA}
  \country{USA}
}
\email{blongboyuan@gmail.com}

\author{Yueqi Wang}
\authornotemark[1]
\affiliation{%
  \institution{Google}
  \city{New York}
  \state{NY}
  \country{USA}
}
\email{yueqiw@google.com}

\author{Hiloni Mehta}
\affiliation{%
  \institution{Google}
  \city{Mountain View}
  \state{CA}
  \country{USA}
}
\email{hiloni@google.com}

\author{Mick Zomnir}
\affiliation{%
  \institution{Google}
  \city{Mountain View}
  \state{CA}
  \country{USA}
}
\email{mzomnir@google.com}

\author{Omkar Pathak}
\affiliation{%
  \institution{Google}
  \city{Mountain View}
  \state{CA}
  \country{USA}
}
\email{omkarpathak@google.com}

\author{Changping Meng}
\affiliation{%
  \institution{Google}
  \city{Mountain View}
  \state{CA}
  \country{USA}
}
\email{changping@google.com}

\author{Ruolin Jia}
\affiliation{%
  \institution{Google}
  \city{Mountain View}
  \state{CA}
  \country{USA}
}
\email{ruolin@google.com}

\author{Yajun Peng}
\affiliation{%
  \institution{Google}
  \city{Mountain View}
  \state{CA}
  \country{USA}
}
\email{yajunpeng@google.com}

\author{Dapeng Hong}
\affiliation{%
  \institution{Google}
  \city{Mountain View}
  \state{CA}
  \country{USA}
}
\email{dapengh@google.com}

\author{Xia Wu} 
\affiliation{%
  \institution{Google}
  \city{Mountain View}
  \state{CA}
  \country{USA}
}
\email{xiawu@google.com}

\author{Mingyan Gao}
\affiliation{%
  \institution{Google}
  \city{Mountain View}
  \state{CA}
  \country{USA}
}
\email{mingyan@google.com}

\author{Onkar Dalal}
\affiliation{%
  \institution{Google}
  \city{Mountain View}
  \state{CA}
  \country{USA}
}
\email{onkardalal@google.com}

\author{Ningren Han}
\affiliation{%
  \institution{Google}
  \city{Mountain View}
  \state{CA}
  \country{USA}
}
\email{peterhan@google.com}
\settopmatter{authorsperrow=4}

%%
%% By default, the full list of authors will be used in the page
%% headers. Often, this list is too long, and will overlap
%% other information printed in the page headers. This command allows
%% the author to define a more concise list
%% of authors' names for this purpose.
\renewcommand{\shortauthors}{Long et al.}

%%
%% This command processes the author and affiliation and title
%% information and builds the first part of the formatted document.

\begin{abstract}
This paper presents a case study of deploying Large Language Models (LLMs) as an advanced "annotation" mechanism to achieve nuanced content understanding (e.g., discerning content "vibe") at scale within an industrial short-form video recommendation system. Traditional machine learning classifiers for content understanding face protracted development cycles and a lack of deep, nuanced comprehension. The "LLM-as-annotators" approach addresses these by significantly shortening development times and enabling the annotation of subtle attributes. This work details an end-to-end workflow encompassing: (1) iterative definition and robust evaluation of target attributes, refined by offline metrics and online A/B testing; (2) scalable offline bulk annotation of video corpora using LLMs with multimodal features, optimized inference, and knowledge distillation for broad application; and (3) integration of these rich annotations into the online recommendation serving system, for example, through personalized restricted retrieval. Experimental results demonstrate the efficacy of this approach, with LLMs outperforming human raters in offline annotation quality for nuanced attributes and yielding significant improvements in user participation and satisfied consumption in online A/B tests. The study provides insights into designing and scaling production-level LLM pipelines for content annotation, highlighting the adaptability and benefits of LLM-based multimodal content understanding for enhancing video discovery, user satisfaction, and the overall effectiveness of modern recommendation systems.
\end{abstract}




\maketitle
\section{Introduction}

High-quality content understanding is crucial for addressing feedback loop challenges in recommendation systems, a task for which traditional Machine Learning (ML) classifiers have been a common approach \cite{Lops2019-eh}. However, they face two primary problems: protracted development cycles for new classifiers, and a tendency towards only high-level content understanding, which can miss nuances crucial for personalized recommendations \cite{deldjoo2024review}. The sheer scale of content on platforms like YouTube further amplifies these issues.

Large Language Models (LLMs) \cite{team2023gemini,team2024gemini, openai2024gpt4technicalreport} offer a transformative solution. This paper investigates the "LLM-as-annotators" approach \cite{2412.05579}, where LLMs serve as human-level content annotators. This can shorten development timelines and enable the annotation of nuanced attributes (e.g., "vibes" like authentic, inspiring, calming, energetic) by leveraging LLM's world knowledge and reasoning.

Applying this in industrial-scale recommender systems presents practical challenges: (1) designing and refining attributes for subtle but important content characteristics; (2) scaling annotation for millions of videos per day with high quality, cost-effectiveness, and low latency; (3) integrating rich annotations into online recommender models to improve user experience and platform metrics.

This paper presents an in-depth case study of deploying an "LLM-as-annotators" system on a large-scale industrial short-form video platform. We detail our end-to-end workflow and share practical learnings on generating nuanced content evaluations and scaling these insights. Our main contributions are:
\begin{enumerate}
    \item Actionable insights into designing and scaling a production LLM annotation system, including its scalable offline inference pipeline and its integration into online recommender systems to improve personalization.
    \item An empirical analysis of the framework's operational challenges and adaptability, detailing strategies to maintain relevance in evolving content landscapes.
\end{enumerate}

This work demonstrates, through a real-world case study, how LLM-driven annotations improve content discovery and user satisfaction, offering practical learnings for leveraging LLMs in recommendation systems. This paper is structured as follows: Section~\ref{sec:methodology} details our methodology, challenges, insights and learnings. Section~\ref{sec:experiment_results} presents experimental results. Section~\ref{sec:conclusion} concludes.

\section{LLM Annotation In Practice}
\label{sec:methodology}

Our methodology for LLM annotation in practice is an end-to-end workflow designed to achieve a deep, nuanced understanding of the vast and dynamic content within large-scale short-form video recommendation systems, moving beyond superficial metadata. This process is segmented into three core stages: (1) defining target attributes and establishing robust evaluation frameworks; (2) offline bulk annotation of content using LLMs and scaling these efforts efficiently; (3) the integration of rich annotations into the online recommender serving system to directly improve user experience. Each stage presents unique challenges related to capturing subtle multimodal characteristics, adapting to rapidly evolving content trends, 
%generating truly actionable insights, 
and ensuring efficient, scalable integration into production. We address these challenges through iterative solutions, leveraging advanced models like Google's Gemini \cite{team2024gemini}, and continuous learning from both offline analysis and online performance.

\begin{figure}[htbp] % The [htbp] are placement specifiers (here, top, bottom, page)
    \centering % This centers the image on the page
    \includegraphics[width=\columnwidth]{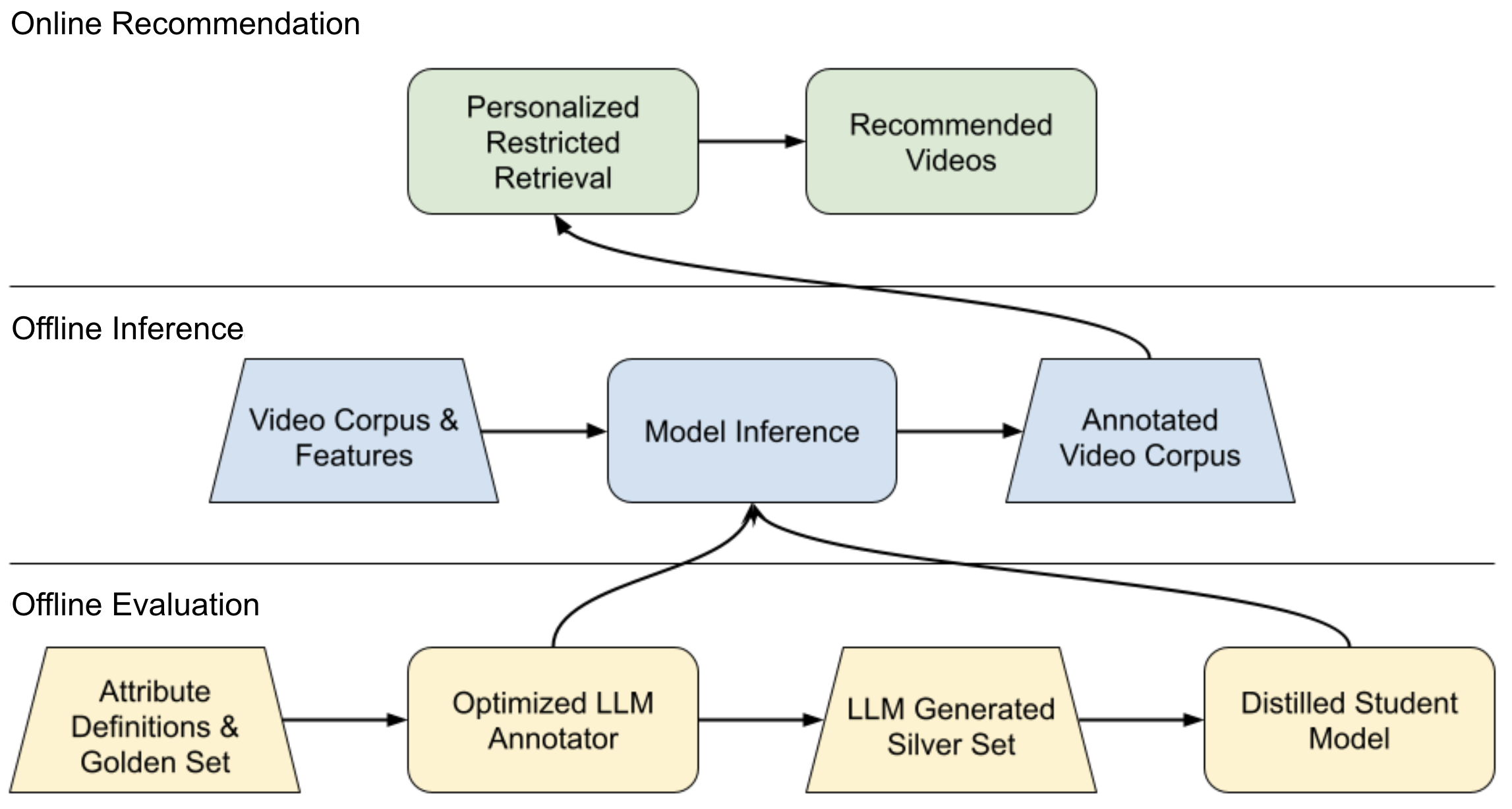}
    % \vspace{-1.5cm}
    \caption{Workflow Diagram. The workflow consists of three stages. Offline Evaluation (yellow) assesses LLM quality using a "Golden Set" of examples validated by aligned expert raters, and this LLM is then used to create a larger "Silver Set" for distilling student models. Offline Inference (blue) processes video corpus and features to build the LLM-annotated corpus. Online Recommendation (green) then leverages these annotations to enhance recommendations through systems like personalized retrieval. Rounded rectangles are models/systems; trapezoids are datasets/processes.}
    \label{fig:recsys} % Optional: for cross-referencing the figure in your text
\end{figure}

\subsection{Defining Target Attributes \& Evaluation}
\label{subsec:defining_evaluation}

The foundation of impactful LLM-driven annotations lies in accurately defining the attributes to be captured and rigorously evaluating the quality of these annotations \cite{ganguli2023challenges}. Thus, this initial stage is crucial for ensuring that LLM's capabilities are directed towards generating meaningful insights for recommendation quality.

Key challenges in this initial stage include developing clear, unambiguous definitions for nuanced or subjective content attributes (e.g., "vibes" like  authentic, inspiring, calming, energetic) that can be consistently interpreted by both LLMs and human evaluators. Furthermore, it's crucial to ensure that LLM-generated annotations are not only accurate according to these definitions but also practically useful and impactful for downstream recommendation tasks. Finally, maintaining the quality and relevance of attribute definitions and evaluation sets as content trends and user expectations evolve presents an ongoing difficulty.

To address these challenges, our approach emphasizes multiple iterations. We start by developing explicit and unambiguous definitions for target attributes. This is supported by a high-quality, manually annotated "Golden Set", created by internal expert raters who achieve alignment through an iterative process of discussions and calibration. This internal alignment is crucial for (1) driving consistent ratings (i.e. improving Inter-Rater Reliability) and (2) refining attribute definition by surfacing edge cases, thereby reducing LLM ambiguity. For example, an initial prompt for the "authentic" vibe might incorrectly exclude some vlog-style videos due to heavier editing/stitching than other "authentic" videos. Reviewing such false negatives allows us to refine the definition, instructing the LLM to prioritize the creator's genuine presentation over the level of editing to better capture the attribute's nuance. Overall, the attribute definition and "Golden Set" annotations provide a foundation for robust offline assessment of LLM performance.

Following this, we employ a careful and iterative strategy for evaluating quality. This strategy involves two key parts: first, assessing performance using offline metrics against our "Golden Set", and second — very importantly — using a continuous improvement cycle based on the results of online A/B testing, which provides direct feedback from real-world deployment. This combined feedback from both offline and online metrics allows us to progressively improve our prompting techniques and make attribute definitions more precise, ensuring that they stay aligned with changing content characteristics and how users engage with the content.

Key learnings from this stage: 
\begin{itemize}
    \item LLM annotation quality for nuanced attributes strongly depends on clarity from internal rater alignment, as reducing human ambiguity directly mitigates LLM ambiguity and improves rating consistency. 
    \item Definitions and LLM annotations significantly benefit from an iterative refinement loop incorporating online A/B testing feedback, with real-world impact as the ultimate quality measure. 
    \item Maximizing the velocity of offline-to-online experiment cycle (including definition refinement, rater calibration, annotation generation, online experimentation, and adjustments) is paramount for achieving highly impactful annotations.
\end{itemize}

\subsection{Offline Bulk Annotation}
\label{subsec:offline_bulk_annotation}

Offline bulk annotation is the step of scaling attribute annotations to a large number of videos. This begins with high-quality LLM annotations on a prioritized set of candidate videos (new, trending, high-impact, etc.) on the order of $10^5-10^6$ videos per day, with sampled video frames, video description, and prompt instructions as input. To make this initial annotation cost-effective, inference efficiency optimization is crucial. Our methodology employs techniques such as model quantization \cite{frantar2023gptqaccurateposttrainingquantization}, batch size tuning, and model sharding optimization \cite{shoeybi2020megatronlmtrainingmultibillionparameter}, yielding a 2-3x throughput increase compared to baseline.

Following the optimized LLM inference, the next step is to extend annotation coverage to the entire video corpus ($O(10^7)$ annotations per day). Our primary strategy to address this broader scaling is knowledge distillation, which enables a significant increase in the quantity of annotations with only minor quality loss. In knowledge distillation, LLM-generated annotations and the probability score for each annotation serve as high-quality teacher labels (i.e. "Silver Set") for lightweight "student" Deep Neural Networks (DNNs). These student DNNs are typically much smaller than the teacher LLM model and trained from scratch on millions of teacher-derived examples using compact features like pre-computed video embeddings. The student models are trained to replicate the teacher's predictive behavior on a given attribute, allowing for annotation to be scaled massively with significantly reduced latency, cost, and only minor quality loss. After passing offline fidelity evaluations, these scaled annotations serve as crucial signals for downstream recommendation models, improving content understanding and personalization.

When working on a new content attribute, we typically start with direct LLM annotations on a prioritized subset of the corpus to evaluate the  headroom through fast iterations. After online A/B testing experiments show positive product impact of an LLM-derived attribute, we further scale the annotation to the entire corpus using knowledge distillation.

Key insights from this offline annotation workflow, particularly regarding inference optimization and knowledge distillation, are:
\begin{itemize}
    \item Inference optimization techniques (model quantization, batch tuning, and others) are critical for the feasibility of using large LLMs, even for teacher labels, offering substantial throughput gains and making initial high-quality annotation more tractable and cost-effective.
    \item Knowledge distillation is a powerful and necessary technique for scaling validated, high-quality LLM annotations to the entire corpus. Training smaller student DNNs on compact features using millions of LLM-derived examples achieves massive annotation throughput at a fraction of the cost and latency of direct LLM inference, democratizing access to nuanced understanding.
    \item The iterative workflow — initial LLM annotation on selected candidates, impact validation, then scaled distillation — provides a robust pathway. It allows agile exploration of novel attributes with powerful models while ensuring only impactful and cost-efficient solutions are broadly deployed.
\end{itemize}

\subsection{Online Personalized Recommendation}
\label{subsec:online_recommendation_serving}

The final stage of our LLM annotation pipeline integrates offline-generated annotations (Section~\ref{subsec:offline_bulk_annotation}) into the online recommendation stack. This translates nuanced content understanding from the annotated corpus into improved real-time video recommendations, aiming for more relevant, engaging, and personalized user experiences that impact key platform metrics.

To integrate the annotations into online recommendation, we used the approach of "Personalized Restricted Retrieval", which allows personalized recommendation of content belonging to specific attributes based on user intent. 
This approach augments candidate generation by performing restrictive nearest neighbor search (e.g., SCANN) \cite{guo2020acceleratinglargescaleinferenceanisotropic} from each attribute's content vocabulary within a production large-scale Transformer-based sequential retrieval model. The triggering of each content attribute is governed by user intent models or heuristics that predict the user's affinity to a given attribute, and the selection of which items to recommend within an attribute is based on the sequential retrieval model's comprehensive understanding of user interests. To enable fast iterations, we built a tight integration of the LLM-annotated corpus with the Personalized Restricted Retrieval stack, which reduced the time from developing a new prompt to online A/B experiments to within one week. In addition, we are also experimenting with alternative approaches, such as adding annotations as features to ranking models or using them as dimensions for diversity treatments.

The impact of online personalized recommendations is measured by continuous A/B testing on key metrics (satisfaction, discovery, ecosystem health). The online experiment insights refine attribute definition (Section~\ref{subsec:defining_evaluation}) and offline annotation (Section~\ref{subsec:offline_bulk_annotation}) through iterative development cycles, and inform launch decisions.

\section{Experimental Results: A Case Study in Production}
\label{sec:experiment_results}

This section presents the empirical validation of our "LLM-as-annotators" workflow within a large-scale, production short-form video recommender system. The evaluation assessed offline annotation quality metrics (Precision, Recall, F1-Score) and online user experience metrics (through A/B experiments), showcasing the production-readiness and practical benefits of this approach.

\subsection{Offline Annotation Quality}
\label{subsec:offline_quality_cs}
We measured Google Gemini's capability for high-fidelity annotation of nuanced content attributes in offline benchmarks. For example, Gemini 2.5 Pro achieved an F1-Score of $81.33\%$ (Precision: $85.03\%$, Recall: $77.94\%$) on a nuanced content attribute, whereas external crowd-sourced human raters\footnote{Human raters were sourced via third-party vendors, aiming for the best feasible annotation quality under project budget and timeline.} achieved an F1-Score of $63.21\%$ (Precision: $76.82\%$, Recall: $53.69\%$) for the same task. This is a critical improvement because the human-rater labels determines the performance upper bound of traditional machine learning classifiers. A key operational learning is that state-of-the-art multimodal Gemini models with flexible prompting greatly improve operational efficiency by eliminating the need for per-task fine-tuning. Together, the human-level annotation quality, iteration efficiency, and tight integration with the online recommender system are crucial for accelerating real-world impact on large-scale production systems.

\subsection{Online A/B Experimentation}
\label{subsec:online_performance_cs}

We conducted online A/B experiments for several LLM-annotated content attributes in our recommender system using personalized restricted retrieval (Section~\ref{subsec:online_recommendation_serving}) and observed statistically significant improvements in key user experience and satisfaction metrics. For example, the experiment for one of the attributes yielded a $+0.49\%$ lift of user participation in content creation, and a $+0.21\%$ increase of satisfied consumption. These results demonstrate that the LLM-driven annotation approach delivers significant benefits on high-volume industrial recommender ecosystems.

\section{Conclusion}
\label{sec:conclusion}

This case study demonstrates the successful deployment of Large Language Models (LLMs), such as Google's Gemini, for nuanced, large-scale content annotation within a production short-form video recommendation system. The "LLM-as-annotators" approach overcomes key limitations of traditional classifiers by enabling a deeper understanding of subtle content attributes and shortening development cycles through flexible prompting of advanced multimodal LLMs. Key learnings and practical insights include the effectiveness of an iterative workflow that combines offline evaluation with online A/B testing to refine annotation quality, the benefits of a scalable production pipeline using knowledge distillation to expand annotation coverage, and the importance of a tight integration with the online recommender stack to maximize experiment velocity. The value of this method was confirmed through both offline evaluation and online A/B testing metrics, showing significant improvements in user experience and content discovery metrics in a large-scale production environment. Future work will focus on deeper integration of LLM annotations with recommender models, and continuous adaptation to the evolving content landscape.

\section{Speaker Bio}
Boyuan Long and Yueqi Wang are software engineers at Google (YouTube) working on recommender systems.

\bibliographystyle{ACM-Reference-Format}
\bibliography{references.bib}

\end{document}